\documentclass[a4paper,aps,pra,showpacs,preprintnumbers]{revtex4}
\usepackage{graphics,graphicx,epsfig}        % для рисунков
\usepackage[centertags]{amsmath}
\usepackage{amsfonts}
\usepackage{amssymb}                     % для математических формул
\usepackage[cp1251]{inputenc}                % для восприятия кодировки Windows
\usepackage[english]{babel}         % для поддержки переносов в выбранном языке
\usepackage{textcomp}                           % всякие символы
\inputencoding{cp1251}
\usepackage{amsthm,color}
\usepackage{euscript}
\begin{document}

\def\BY{\begin{eqnarray}}
\def\EY{\end{eqnarray}}
\def\BE{\begin{equation}}
\def\EE{\end{equation}}
\def\BEA{\begin{eqnarray}}
\def\EEA{\end{eqnarray}}
\def\k{\mathbf{k}}
\def\q{\mathbf{q}}
\def\r{\vec{r}}
\def\ro{\vec{\rho}}
\def\L{\label}
\def\nn{\nonumber}
\def\({\left (}
\def\){\right )}
\def\[{\left [}
\def\]{\right]}
\def\<{\langle}
\def\>{\rangle}
\def\h{\hat}
\def\hs{\hat{\sigma}}
\def\td{\tilde}
\def\ds{\displaystyle}

%-----------------------------------------------------------
\title{Preservation of quantum correlations in femtosecond light pulse train  within  atomic ensemble } \vspace{1cm}
\author{ {A.D.~Manukhova}, {K.S.~Tikhonov}, {T.Yu.~Golubeva}, {Yu.M.~Golubev} }
\address{Saint Petersburg State University, 198504 St.Petersburg, Petershof, Ulyanoskaya st. 1, Russia}
\date{\today}

\begin{abstract}

In this paper, we examined a possibility of preservation of a substantially multimode radiation in a single cell of quantum memory.
As a light source we considered a synchronously pumped optical parametric oscillator (SPOPO).
As it was shown in \cite{Patera,Roslund,Araujo,Gerke},  SPOPO radiation has a large number of the correlated modes making it attractive for the purposes of the quantum communication and computing.
We showed that these correlations can be mapped on the longitudinal spin waves of the memory cell and be restored  later in the readout light.
The efficiencies of the writing and readout depend on the mode structure of the memory determined by a mechanism of the light-matter interaction under consideration
(the non-resonance Raman interaction) and by the profile of the driving light field.
We showed that like the initial light pulse  train, the restored one can be represented by a set of  squeezed supermodes.
The mapping of the quantum multimode correlations on the material medium offers opportunities to manipulate the quantum states within the memory cell followed by the reading of the transformed state.

\end{abstract}
 \maketitle

%-----------------------------------------------------------

\section{Introduction}

Nowadays, study of various quantum memory mechanisms is one of the key challenges in quantum communication over large distances, as well as quantum computing.
Memory cell, that allows one to receive quantum signal on demand, is a principal element of a quantum repeater  \cite{Quantum repeater Mazurek, Quantum repeater Li}
and it is included in a scheme of quantum logic transformations \cite{Fredkin Gate, Toffoli Gate,Simon2010,Zhong2015}.
At present, a number of schemes imprinting  quantum state of the field on the long-lived degrees of freedom available in  the material medium has been found.
They are based on different mechanisms of interaction: effect of Electromagnetically Induced Transparency 
 \cite{Fleischhauer2005,Liu2001,Phillips2001,Novikova2012, EIT Choi, EIT Novikova}, Quantum Nondemolition schemes \cite{Muschik2006, Julsgaard2004, QND Zhang, QND
Lliang}, Raman interaction in a $\Lambda$-configuration  \cite{Nunn2007, Samburskaya2011, Gorshkov200776, Reim2010, Quant},
 high-speed resonant memory \cite{Golubeva2011, Tikhonov2014}, Controlled Reversible Inhomogeneous Broadening \cite{Moiseev2007, Sangouard2007, CRIB Longdell, Iakoupov2013},  
 and  Atomic Frequency Comb \cite{AFCAfzelius,Jobez2014,Saglamyurek2014, Bonarota2011, AFC Gerasimov}.

However,  recently a memory cell becomes considered not only as a "container"\; for a quantum signal, but also as a microchip (quantum transistor),
which allows not only to preserve but also to manipulate with signals inside  \cite{Kuzmin,Moiseev,Chen}.
In that respect, we would like to examine the possibility of implementing such a memory cell to perform quantum calculations directly within the cell.

For this purpose, we need to preserve several modes within one memory cell. Besides, these modes have to be not independent but to have quantum correlations between them.
This essentially distinguishes the problem considered here from the previous studies.We regard the preservation of the cluster state of light and manipulation with it within a single memory cell as a strategic option.

To achieve this goal, we need to choose a source of multimode light with strong entanglement.
A good candidate for this is a radiation of synchronously pumped optical parametric oscillator (SPOPO) \cite{Patera}. SPOPO is a high-Q ring cavity with a parametric crystal inside.
The cavity is pumped by short periodic pulses of light. The duration of each pulse is of about 100 femtoseconds. When the pump pulse passes through the crystal, correlated pairs of signal photons are created.
The main feature of SPOPO is that the repetition period of the pump pulses coincides with resonator round-trip time both of the signal and  pump fields.
Thus, every time the fields pass through the crystal, they are involved in the generation process of the next photon pairs.This creates additional correlations in the system.

Such a light has been experimentally obtained in \cite{Roslund,Araujo}. Examining quantum correlation of this light, the authors showed that the
SPOPO radiation reveals genuine multipartite entanglement. They observed experimentally six so-called "supermodes": six independent quantum channels
(q-modes) which are subsumed within the comb. Let us note that theoretically even higher number (about 100) of squeezed modes was predicted to be
embedded in the frequency structure of the quantum comb. Moreover, the authors showed that it is possible to fabricate the cluster structures
necessary for computation from the modes contained within the quantum comb. All aforesaid makes this light extremely attractive for quantum
computation.

In this article, we will show that quantum properties of SPOPO can be effectively preserved in a single memory cell.
However, direct writing and readout do not preserve the inherent correlation properties of the signal field. To get over this difficulty,  phase-shifts of  input and output signals must be implemented.
In this case, it is possible not only to obtain high efficiency of a multimode storage, but also to preserve the quantum correlation.
As a mapping mechanism of the quantum state of the field to the medium, we have chosen  the Raman scheme of interaction of fields with atomic ensemble in a $\Lambda$-configuration.

The article is structured as follows. In Section \ref{II}, we build a model of train of ultrashot light pulses coupled with an atomic ensemble in the presence of the driving field in  the Raman configuration. 
Basic approximations are also provided here. In Section \ref{III}, the set of  Heisenberg equations is solved at the writing and the readout, and the possibility of applying  
the Schmidt decomposition for further analysis is justified. In Section \ref{source}, 
we consider SPOPO radiation as a source of the signal field in the quantum memory scheme and formulate quantum statistical characteristics of this radiation. 
An estimation of the writing efficiency and a preservation of the quantum correlations of the input signal at the readout are presented in Sections \ref{record} and \ref{reading}, respectively. 
Section \ref{VII}  is dedicated to an examination of the SPOPO squeezed supermodes storage.

%----------------------------------------------------------------------------

\section{Quantum memory model based on the multifrequency comb}\L{II}

Let us consider the model of storage of the correlated femtosecond light pulse train in a motionless atomic ensemble with the  $\Lambda$-type energy level configuration.

\begin{figure}[h]
 \begin{center}
  \includegraphics[height=45mm]{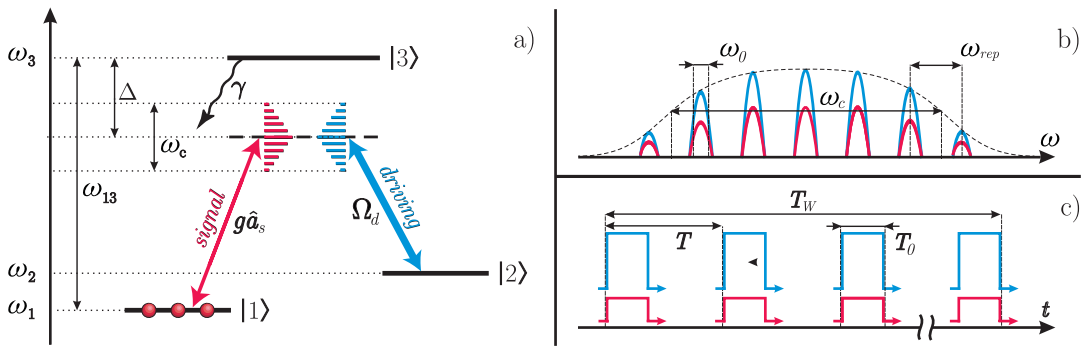}
\caption{a) Energy level scheme of an atomic ensemble, interacting with combs of signal and driving in the Raman memory protocol;
 b) frequency profile of the signal field:   $\omega_c$ -- spectral width of the comb, $\omega_{rep}$ -- distances between the comb's teeth,
$\omega_0$ --  broadening of a single tooth; c) time profile of the signal field: $T_W$ --  total duration of the periodic pulse train, $T$  --
repetition period, $T_0$  --  duration of a single pulse.}
  \label{Raman1}
 \end{center}
 \end{figure}

Figure \ref{Raman1} shows the relevant atom-field interaction scheme.
Three-level cold atomic ensemble of length $L$  elongated in the $z$  direction interacts with the quantum signal field $\hat{\vec{E}}_s$ in the presence of the strong classical driving field $\vec{E}_d$.
Carrier frequencies $\omega_s$ and  $\omega_d$ of signal  and driving are detuned from the resonance frequencies $\omega_{13}$ and $\omega_{23}$ of the atomic transitions, respectively
($\Delta=\omega_s-\omega_{13}=\omega_d-\omega_{23}$).

Initially all atoms are prepared in the ground states $|1 \rangle$  using optical pumping. Field polarizations are adjusted so that the signal field acts on the transition between the levels  $|1  \rangle$ and $|3  \rangle$, 
and the driving one acts  on the $|2  \rangle$ - $|3  \rangle$  transition. We assume level $| 3 \rangle$  to be  short-lived with the lifetime $\gamma^{-1}$. 
The decay of  lower levels $| 1 \rangle$ and $| 2 \rangle$  is neglected since we consider these levels to be long-lived compared with the memory storage time.

We solve the one-dimensional spatial problem and regard the driving field as a strong classical plane wave and the signal field as a weak quantum one.
In this paper, we consider as a signal field a frequency comb generated in the process of spontaneous parametric down-conversion of light \cite{Fabre},
the structure of this field will be described in more detail in Section \ref{source}. Both fields represent ultrashort pulse trains and propagate  in positive direction of the $z$ axis.
The pulses of the signal field are correlated within the times of the order of cavity photon lifetime. 
In the frequency representation such fields have the form of the frequency comb with the spectral width  $\omega_c$ (see. Fig. \ref{Raman1}), teeth of which are broadened depending on the number of pulses in the train. 
In this model, we assume $\omega_c \ll |\Delta|$ that allows us to consider the detuning for each  tooth of the comb  to be the same.
We choose the single pulse duration $T_0$ and repetition period $T$ to be the same for the driving and signal fields. We assume that $T_0 \ll L/c \ll T$, so that at any given time the medium contains not more than one pulse. 
Note that all these relations  between the parameters are fully consistent with experimental data \cite{Fabre}. Thus, analytical expressions for signal and driving fields can be written as:
\BY
&& \hat{\vec{E}}_s(t,z)=-i\; \vec{e}_s \; \sqrt{\frac{\hbar\omega_s}{2\varepsilon_0 c}} \;  \hat{a}(t,z)e^{-i\omega_st+ik_sz} +h.c.\; ,\L{1}\\
&& \vec{E}_d(t,z)=-i \; \vec{e}_d  \; E_0(t,z) e^{-i\omega_dt+ik_dz}+h.c.\L{2}
\EY
Here $k_ {s, d}$ are the wave numbers, $ \vec{e}_ {s,d} $ are the polarization vectors of the signal and driving fields, respectively, $ E_0 (t, z) $ is 
the classical control field amplitude, $\hat{a}(t,z)$ -- annihilation bosonic operator of the slowly varying amplitude of the signal field, which obeys the following commutation relations:
\BY
&& [\hat{a}(t,z),\hat{a}^\dag(t,z')]=c\(1-\frac{i}{k_s}\frac{\partial}{\partial z}\)\delta(z-z'),\L{3}\\
&& [\hat{a}(t,z),\hat{a}^\dag (t',z)]=\delta(t-t').\L{4}
\EY
To describe the atomic medium, we introduce the collective operators of coherences and populations:
\BY
&& \hat{\sigma}_{ij}(t,z)=\sum_{k=1}^{N_{at}}\hat{\zeta}_{ij}^k(t)\delta (z-z_k),\L{5}\\
&& \hat{N}_{i}(t,z)=\hat{\sigma}_{ii}(t,z)=\sum_{k=1}^{N_{at}}\hat{\zeta}_{ii}^k(t)\delta (z-z_k),\L{6}
\EY
where $ N_{at} $ is the number of atoms in the ensemble, $ \hat {\zeta} _ {ij}^k (t) $ is the spin coherence between  $ | i
\rangle$ and $ | j \rangle $ levels of the $ k $\!--th atom
at  time $ t $, $ \hat {\zeta}_ {ii}^k (t) $ is the population of $| i \rangle$ level of the $k$\!--th atom at time $ t $. It is easy to see that these operators obey the following quantum permutations:
\BY
[\hat{\sigma}_{ij}(t,z),\hat{\sigma}_{ji}(t,z')]=\(\hat{\sigma}_{ii}(t,z)-\hat{\sigma}_{jj}(t,z)\)\delta(z-z').\L{7}
\EY
Then  the interaction Hamiltonian in the rotating wave and dipole approximations can be derived in form:
\BY
 \hat{V}=&&\int d z  (i\hbar g (\hat{\sigma}_{31}(t,z)\hat{a}(t,z)e^{-i\Delta t+ik_sz}- \hat{\sigma}_{13}(t,z)\hat{a}^\dag(t,z)e^{i\Delta t-ik_sz})
\L{8}\\
&&+ i\hbar\Omega(t,z)(\hat{\sigma}_{32}(t,z) e^{-i\Delta t+ik_dz}-\hat{\sigma}_{23}(t,z) e^{i\Delta t-ik_dz})),\nn
\EY
where $g=\sqrt{\omega_s/2\varepsilon_0 c \hbar}\;d_{13}$ is the coupling constant  between a single atom and the signal field,
 $\Omega(t,z)=d_{23}E_0(t,z)/\hbar$ is the Rabi frequency of the $|3\rangle$ -- $|2\rangle$ transition induced by the driving field;
 $d_{ij}$ are the matrix elements of the dipole moment operator corresponding to the  $|i \rangle$  --  $|j \rangle$ level transitions, which for simplicity are assumed to be real.

Since the signal field is generated by the SPOPO, the characteristic signal field profile
will be determined by the laser parameters  \cite{Fabre}. However, without limiting the generality, we can talk about preservation of the quantum correlations in the stochastic radiation with a rectangular profile. 
In general, the problem can be solved by  choosing the driving field profile ensuring preservation of the required statistical characteristics of the signal. Here, for simplicity we consider  driving profile be identical to the profile of the signal.
We treat the driving field as a plane wave traveling in the positive direction of the z axis. This wave is defined by the Rabi frequency $\Omega(t,z)$. Because of the wavefront delay by the value $ z / c $,
expression for the Rabi frequency could be written as $ \Omega (t, z) = \Omega_0F (t-z / c) $, where the function $ F (t-z / c) $  describes  time profile of the driving during  the writing and readout:
\BY
&&\Omega(t,z)=\Omega_0F(t-z/c),\qquad F(t)=\sum_{n=1}^N\Theta(t-t_n),\qquad t_n=(n-1)T, \L{9} \\
&& \Theta(t)=H(t)\cdot H(T_0-t).\nn
\EY
Let us remind that $ N $ is the total number of pulses, $T_0$ is the  duration of each single pulse, and $T_0\ll L/c $.  The repetition period of the train $ T\gg
T_0,\;L/c$. Every $n$-th pulse enters the medium at time $t_n=(n-1)T$ ($n=1,2,\cdots,N$), and $H(t)$ is the Heaviside function.

Thus, each pulse of the driving field corresponds to the appropriate signal one. Note that, according to the definition~(\ref{9}),  $\Omega^2(t,z)=\Omega_0^2F(t-z/c)$.

We want to emphasize once more the important feature of the problem to be solved here: 
as distinct from the single pulse storage \cite{Nunn2007},
here we consider the preservation of the pulse sequence, which carries the complex multi-mode quantum state, 
and each pulse enters the medium already modified by the previous ones.
Moreover, the aim is to preserve not each separate pulse in the medium, but quantum correlations between them.

Taking into account the commutation relations (\ref{3})~-~(\ref{4}) and (\ref{7}) we can easily obtain the system of Heisenberg equations for the slowly varying amplitudes. 
According to Wigner-Weisskopf theory, the resulting system should be supplemented by relaxation terms and corresponding Langevin noise sources, which describe the spontaneous decay from the level  $ | 3 \rangle $. 
Then the system is averaged over the atomic positions. 
Assuming that the change of population of the level $ | 1\rangle $  in the memory process is negligible, we can replace the difference $\overline{\hat{N}}_1-\overline{\hat{N}}_3$   by the  $c$-number $N_{at}/L$, 
which characterizes   linear atomic density. Eliminating adiabatically  the level  $ | 3 \rangle $, we obtain the closed set of two equations:
\BY
&&\(\frac{\partial}{\partial t}+c\frac{\partial}{\partial z}+icB\)\hat{a}(t,z)=-icCF(t-z/c)\(\hat{b}(t,z)+\hat{f}_a(t,z)\),\L{10} \\
&&\(\frac{\partial}{\partial t}+iAF(t-z/c)\)\hat{b}(t,z) = -iCF(t-z/c)\(\hat{a}(t,z)+\hat{f}_b(t,z)\),\L{11}
\EY
where  $\hat{b}(t,z)=\hat{\sigma}_{12}(t,z)/\sqrt{N_{at}/L}$\; is the spin coherence operator renormalized to satisfy
bosonic commutation relations;   $\hat{f}_a(t,z)$  and $\hat{f}_b(t,z)$ are the Langevin noise sources. Considering the Raman limit  $|\Delta|\gg\gamma$,  factors $ A $, $ B $, and $ C $  can be simplified by the way:
\BY
&&A=\frac{-i\Omega_0^2}{\gamma/2-i\Delta} \to \frac{\Omega_0^2}{\Delta},\L{12}\\
&&B=\frac{-ig^2N_{at}/L}{\gamma/2-i\Delta}  \to  \frac{g^2N_{at}/L}{\Delta},\L{13}\\
&&C=\frac{-i g\sqrt {N_{at}/L}\;\Omega_0}{\gamma/2-i\Delta}  \to \frac{g\sqrt {N_{at}/L}\;\Omega_0}{\Delta}\;.\L{14}
\EY
By the nature of the case, as we solve the problem in the time representation, we take into account not only  exact two-frequency resonance for each tooth of comb but also all nonresonant terms within the width of the comb.
Such an approach distinguishes this consideration from that of \cite{Oxana}. Looking ahead, let us say that other feature is, that storage of the signal field with a complex structure in the memory cell in addition to traditional characteristics of quantum memory  allows to investigate questions about preservation of quantum correlations, contained in the input signal, and their imprinting on  mode structure of the memory cell.

\section{Solution of Langevin-Heisenberg equations. Schmidt Modes}\L{III}

Equations   (\ref{10})~-~(\ref{11}) describe evolution of the field and material variables during the time of light-atom interaction. Depending on the chosen initial conditions, times of these interactions
may correspond to the writing or the readout of the signal.

The process of the writing of the quantum state of light on the atomic ensemble suggests that the signal field $\hat{a}_{in}(t)$ enters the front face of the memory cell ($ z=0 $),
and, besides, spin oscillator of the atomic subsystem is initially in the vacuum state. With these initial and boundary conditions, we solve the set (\ref{10})~-~(\ref{11}), using Laplace transform method and obtain the expression for the spin coherence:
\BY
&&\hat {b}(t,z)=-iCe^{ -iBz}\int\limits_{0}^{t-z/c} dt^\prime F(t^\prime)\;\hat a_{in}(t^\prime)\;e^{ -iAQ(t-z/c,t^\prime)}
\;J_0\(2C\sqrt{Q(t-z/c,t^\prime)z}\)+vac,\L{22}
\EY
where
\BY
&& Q(\eta,\eta^\prime)=\int\limits_{\eta^\prime}^{\eta}d\eta^{\prime\prime}F(\eta^{\prime\prime}).\L{16}
\EY
With $vac$ hereafter we mean the contributions from the vacuum channels.
To analyze the coherence formed in the writing process, we would only be concerned with the solution that occurs when the comb has passed through the medium.
The writing of the train of pulses at every point of the medium lasts
 $T_W=t_N+T_0$. Defining $\hat {b}(z, t=t_N+T_0)=\hat {b}^W(z)$ and taking into account the explicit expression of the function $F(t)$, we  get the spin coherence by the end of the writing time:
\BY
&&\hat {b}^W(z)=-iCe^{ -iBz}\sum_{n=1}^N\int\limits_{t_n}^{t_n+T_0} dt\;\hat a_{in}(t)\;e^{ -iA(D_n+t_n-t)} \;J_0\(2C\sqrt{(D_n+t_n-t)z}\)+vac,
\L{17}
\EY
where $D_n= (N-n+1)T_0$.\\

It can be shown that the coefficient $ B $ has the meaning of negligibly small additive to the wave number $ k_d $ in the expression for the fast varying amplitude  of the
spin coherence, so one can put $ B = 0 $ (see Appendix \ref{A}):
\BY
&&\hat {b}^W(z)=-iC\sum_{n=1}^N\int\limits_{0}^{T_0} dt\;\hat a_{in}(t+t_n)\;e^{ -iA(D_n+t_n-t)}\;J_0\(2C\sqrt{(D_n-t)z}\) +vac.\L{18}
\EY
Note that we assume the storage process to be ideal, so the spin coherence during the storage time remains the same \cite{Tikhonov2015}.

To describe the readout, we solve Eqs. (\ref{10})~-~(\ref{11})  with other initial and boundary conditions:
quantum mode of the signal field is in a vacuum state, and the spin coherence distribution coincides with that  obtained at the end of the writing.
We consider the time profile $ F(t) $ of the driving during the reading to be the same as during the writing.
Note that generally there are two possibilities of the readout - the forward readout (when the reading driving field has the same direction as the fields at the  writing)
and the backward one (when the writing driving field and reading one are passed in the opposite directions). Further, we will only be concerned with the backward readout as the most effective one \cite{Nunn2007}.
Thus, the expression for the amplitude of the output signal is as follows:
\BY
&&\hat a^R(t)=-iC F(t-L/c)\int\limits_0^L dz^\prime \hat { b}(z^\prime,t=0)\;e^{-iBz^\prime}\; e^{ -iAQ(t-L/c, -L)} J_0\(2C\sqrt{Q(t-L/c,
-L)z^\prime}\)+vac.\L{19}
\EY
Based on the same reasons that are mentioned above, we can put  $B=0$ once again and, taking into account the explicit expression of the function $F(t)$, we get that during the readout the amplitude of the signal is developing as:
\BY
&&\hat{a}^R(t)=-iC\int\limits_0^L dz^\prime \hat { b}^W(z^\prime)\;\sum_{n=1}^N\Theta (t-t_n)\; e^{
-iA((n-1)T_0+t-t_n)}J_0\(2C\sqrt{((n-1)T_0+t-t_n)z^\prime}\)+vac.\L{20}
\EY
Once again let us compare these solutions with those found in \cite{Nunn2007}. Despite the fact that the problems are formulated in a similar manner, the main difference is
that the duration of the pulse to be stored  is deliberately chosen in  \cite{Nunn2007}  to be much shorter than the length of the medium in the time representation ($L/c$), while  in our case, 
in spite of that each single pulse of the train is still much shorter than the medium,  the whole train significantly exceeds the length of the medium $L/c$.
In analyzing the results, this difference is reflected in such memory characteristics as efficiency
and the preservation of the second-order correlator, as it will be discussed below in Section \ref{record}.

For further convenience, all formulas and expressions will be given in dimensionless variables
 $z$ and $t$, so the spatial variable is expressed in units of the optical depth,  scaled to $\gamma/\Delta$ , and the time variable  -- in units of the Rabi frequency, scaled to
$\Omega_0/\Delta$:
\BY
&&\frac{\Omega_0^2}{\Delta}t\rightarrow t \:,\qquad \frac{g^2(N_{at}/L)}{\Delta}z\rightarrow z \;.\L{21}
\EY
Let us select in the Eqs. (\ref{18}) and (\ref{20}) the integral kernels describing the memory process:
\BY
&&\hat{b}^W(z)=-i \int\limits_0^{T_W}dt  \;G_{ab}(T_W-t,z)\hat{a}_{in}(t)+vac. \:,\\
&&\hat{a}^R(t)=-i\int\limits_0^L dz \;G_{ba}(t,z)\hat{b}^W(z)+vac.,\L{23}
\EY
where
\BY
&&G_{ab}(t,z)=\sum\limits_{n=1}^{N}\Theta(t-t_n)e^{ -i(D_n+t_n-t)}J_0(2\sqrt{z((N-n+1)T_0+t_n-t))}),\L{24}\\
&&G_{ba}(t,z)=\sum\limits_{n=1}^{N}\Theta(t-t_n)e^{ -i((n-1)T_0+t-t_n)} J_0(2\sqrt{z((n-1)T_0+t-t_n)}).\L{25}
\EY
Since our further analysis will be concerned only with the averages of the normally ordered quantities, we can omit the contributions of $vac$  in these expressions.

The kernels of the integral transformations $G_{ab}(t,z)$ and $G_{ba}(t,z)$ are linked by the following relationship (see Appendix \ref{B}):
\BY
&&G_{ab}(t,z)=G_{ba}(T_W-t,z).\L{26}
\EY
It is convenient to introduce the kernel $K(t,t')$, that expresses the relationship between the initial signal and the one to be restored after the full cycle of writing, storing and readout:
\BY
&&K(t,t')=\int\limits_0^LG_{ba}(t,z)G_{ba}(t',z)dz.\L{27}
\EY
Then the input and output  signals are connected to each other by integral equation:
\BY
\hat{a}^R(t)=-\int\limits_0^{T_W}dt'K(t,t')\hat{a}_{in}(T_W-t').\L{28}
\EY
To be able to apply the Schmidt modes technique, it is necessary and sufficient  for the kernel of the integral transformation of the full cycle to be Hermitian. 
Let us emphasize that using the Schmidt decomposition for analyzing the memory scheme is very convenient because it explicitly shows the number of independent degrees of freedom that can be preserved in the scheme.
However, the physical requirements to the memory process make us to impose more stringent conditions on the properties of the kernels.
If the integral transformation is characterized by a complex kernel, the quadratures of the signal do not evolve independently. It inevitably leads to a loss of the shot noise suppression in the restored field.
Let us show how to overcome this problem in our case.

The kernel  $K(t,t')$ can be represented as the product of a phase factor comprising imaginary exponent functions and the amplitude part named $G(t,t')$.
Note that $G(t,t')$, the amplitude of the kernel, is real and symmetric with respect to permutations of the arguments, and, therefore, is representable as a Schmidt eigenfunction expansion.
First, let us look at the shape of these eigenfunctions. We will show that they have the comb structure of the pump, and will demonstrate that for the parameters of interest, this structure is negligible
to assess the characteristics of the quantum memory.
Let us demonstrate this on the example of the first eigenfunction of the kernel
$G(t,t')$.

\begin{figure}
 \begin{center}
  \includegraphics[height=65mm]{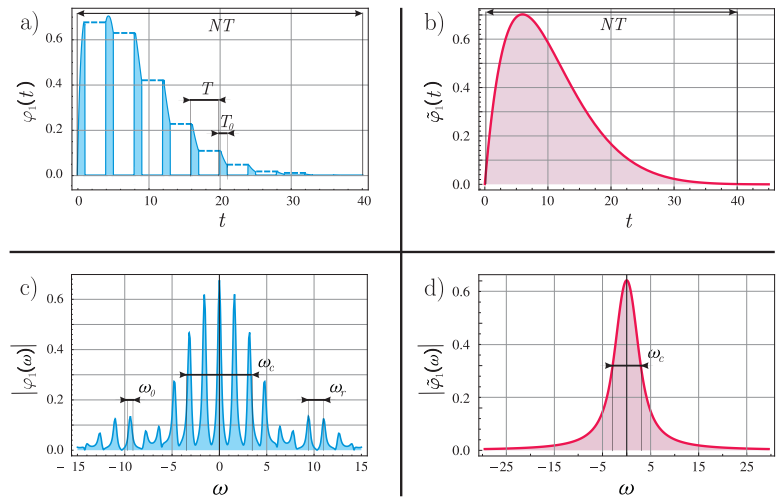}
 \caption{First eigenfunction of the kernel of the full memory cycle and its envelope at time and frequency representations.
Parameters for the calculation: $N=10$,  $T_0=1$, $T=4$, $L=10$. }
  \label{functions}
 \end{center}
\end{figure}

Figure \ref{functions} shows the senior eigenfunction of the kernel $\varphi_1(t)$ (the eigenvalue $\lambda_1=1$), its envelope $\tilde{\varphi}_1(t)$ and their spectra.
To demonstrate visually the comb structure of the functions, the calculation was carried out under the parameters that are far from the real:
we reduced the number of pulses in the train to 10
 (usually this value is about 100), and chose the repetition period $ T $ to be of the same order with the duration of a single
pulse $T_0$ (remind  that for the real comb  $T\gg T_0$, as we will continue to assume further). Approaching parameters to the real ones, i.e., increasing the number of pulses in the train ($ N >> 1 $), the meaningful  time evolution  will be determined only by the envelopes of the eigenfunctions $\tilde{\varphi}_i(t)$, and the evolution on the intervals $T_0$ can be neglected when assessing the operation of the memory cell.

Note that the eigenfunctions $\varphi_i(t)$ of the kernel $ G (t, t ') $ and their envelopes of $\tilde{\varphi}_i(t)$ are normalized to  interval $[0,T_W]$ as follows:
\BY
&&\int\limits_0^{T_W}\varphi_i(t)\varphi_j(t)dt=\delta_{ij} ,\qquad\qquad
\int\limits_0^{T_W}\tilde{\varphi}_i(t)\tilde{\varphi}_j(t)dt=\frac{T}{T_0}\delta_{ij} .\L{29}
\EY

Then the integral transformation of the full cycle can be derived as (see Appendix  \ref{C}):
\BY
&&\hat{a}^R(t)=-\int\limits_0^{T_W}dt' e^{-i   \frac{T_0}{T}   (t+t')}G(t,t') \hat{a}_{in}(T_W-t').\L{30}
\EY
where
\BY
&&G(t,t')=\int\limits_0^L F(t)F(t')J_0\(2\sqrt{ \frac{T_0}{T} zt}\)J_0\(2\sqrt{ \frac{T_0}{T}
zt'}\)dz=\sum\limits_{i}\sqrt{\lambda_i}\;\varphi_i(t)\varphi_i(t')=
\frac{T_0}{T}\sum\limits_{i}\sqrt{\lambda_i}\;\tilde{\varphi}_i(t)\tilde{\varphi}_j(t').\quad\L{31}
\EY

The presence of the exponential factor in the kernel of the full cycle $K(t,t')$ substantially changes the properties of the memory.
In Sections \ref{record}  and  \ref{reading}, we will show that though it is possible to achieve high efficiency of the memory by choosing  the appropriate ratio between the length of the cell and the number of pulses to be stored, the correlation properties of the field  would be destroyed in the memory process.
We will offer a scheme that allows one to compensate for the effect of phase factors and restore the cell operation.

\begin{figure}[h]
 \begin{center}
  \includegraphics[height=55mm]{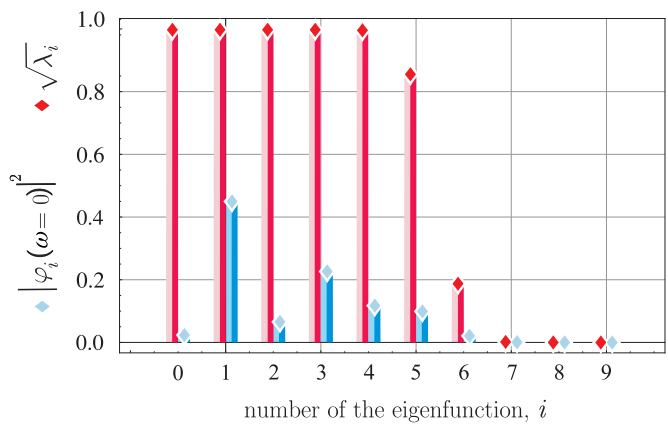}
 \caption{Eigenvalues of the kernel $G(t,t')$ (red bars) and  zero spectral components of the eigenfunctions (blue bars).  Parameters for the calculation: $N=90$,  $L=10$, $T_0=0.1$, $T=10000$.}
  \label{eff}
 \end{center}
\end{figure}

Since the further analysis of the solutions will be performed in terms of the  eigenfunctions and eigenvalues of the kernel  $G(t,t')$, we calculated the eigenvalues for the parameters of the signal, identified  in accordance with the experiment \cite{Araujo}. In Fig. \ref{eff}, the red bars show the eigenvalues of the full  cycle transformation for the chosen parameters of the problem.
We see six eigenvalues close to 1, i.e., our memory,  in principle, is able to preserve not more than six modes for the chosen parameters. To reveal the preservation of what quantum characteristics
of the radiation we should follow in the process of memory, and, in particular, how many independent quantum degrees of freedom SPOPO's radiation has, let us discuss its quantum statistical properties.

%----------------------------------------------------------------------------

\section{Quantum-statistical properties of light generated by SPOPO}\L{source}

As a signal field we consider a frequency comb produced in the process of a degenerate parametric light scattering in a synchronously pumped cavity \cite{Patera}.
The process of the signal comb generation is realized as follows: 
the nonlinear crystal inside the cavity is pumped by a sequence of coherent pulses with a duration of a single one $T_0$.
When the photons of the pump pass through the crystal, pairs of correlated photons of the signal are created.
The cavity round-trip time $T$ of the signal pulse coincides with the repetition period of the pump ($T_0\ll T$).
Taking into account the resonator Q-factor,  photons of the signal can make about ten round-trips over the cavity before leaving.
Thus, each time passing through the crystal, they are involved in the creation of the next pair of the correlated photons.
Therefore, there is a wide variety of correlations in the system at different times.
The study of the correlation properties of the SPOPO's light showed that the correlations within one single pulse are much weaker than the correlations between consecutive pulses of the train.
In  \cite{Averchenko2011}  authors neglected correlations within each single pulse and focused on the correlations of the system at times comparable and longer than the repetition period of pulses.
It was shown in \cite{Gerke} that these correlations are a resource for quantum computing. 
We will take the same approach. It should be noted that this approach is also consistent with the capabilities of our memory scheme,
as it will be discussed below in Section \ref{reading}.

As is well known, SPOPO can operate both in the sub-threshold and  above-threshold regimes  \cite{Averchenko2011,Averchenko2011a}. In this paper, we consider the light received from the SPOPO  below its oscillation threshold.
In this mode it is possible to neglect the depletion of the pump under its parametric conversion in the nonlinear crystal
and we can assume that the pump is in a coherent state with the amplitude specified by the external source. We imply that all pulses in the train have the same shape since the generator operates
 in a steady-state regime.

The amplitude of the signal can be represented as a superposition of the individual pulses:
\BY
&&\hat A(t)=\frac{1}{\sqrt{N}} \sum_{m=1}^N \hat A_m(t-t_m)=\frac{1}{\sqrt{N}} \sum_{m=1}^N \hat {a}_m\Theta(t-t_m),\L{32}
\EY
and obeys the following commutation relations:
\BY
&&\[\hat A(t),\hat A^\dag(t^\prime)\]=\delta(t-t^\prime),\qquad\[\hat A_n(t-t_n),\hat A_m^\dag(t^\prime-t_m)\]=\delta_{mn}\delta(t-t^\prime),\L{33}
\EY
where $N$ is the number of pulses in the train.

Speaking of the quantum statistical properties of the SPOPO, quadrature squeezing that was  observed for the $\h Y$-quadrature should first be noted.

\begin{figure}[h]
 \begin{center}
  \includegraphics[height=40mm]{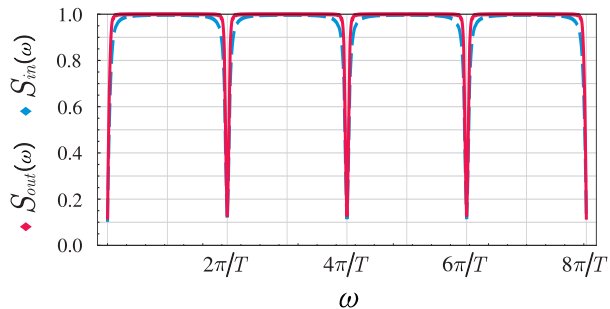}
 \caption{Reduction of the short noise in spectrum of the photocurrent fluctuations under homodyne detection of the $\h Y$-quadrature of $N$ consecutive pulses of the input (blue dashed line) 
 and output (red solid line) signal fields. The time profile of the homodyne $\beta(t)$ is a step function in both cases.
Parameters of the calculation:
 $N=90$, $T_0=0.1$, $T=10000$, $\kappa_sT=0.1$, $L=10$.}
  \label{CorrAfter}
 \end{center}
\end{figure}

We suppose to study the preservation of the quadrature squeezing, basing on the time representation, as it seems to be more natural language for the quantum memory problem.
In \cite{Averchenko2011} the correlation function of the $\h Y$-quadrature fluctuations in different pulses was obtained:
\BY
&& \langle \h Y(t-nT)\h Y(t'-n'T) \rangle=\frac{1}{4}\delta_{nn'}\delta(t-t')\nn\\
&&-\frac{\kappa_sT}{4}\frac{\Theta(t-nT)}{1+\Theta(t-nT)}e^{-\frac{\kappa_sT}{2}(1+\Theta(t-nT))|n-n'|}\;\;
\delta(t-t'-(n-n')T)\Theta(t-nT)\Theta(t'-n'T),\label{34}
\EY
where $n$ and $n'$ are numbers of pulses; $\kappa_s$ is a  spectral width of the resonator; $T$ is the repetition period of the signal;
$\Theta(t)$ is the  time profile of a single pulse, which, for simplicity, we assume to be rectangular since that does not change fundamental properties of the light to  be studied.
The first term in Eq. (\ref{34})  corresponds to the shot noise, the presence of the second, negative one, shows the ability to suppress the shot noise, i.e., it indicates the presence of quantum correlations in the system.
In order to obtain the signal correlator, we need to sum Eq. (\ref{34})  up for all pulses, numbered by the indexes $n$ and $n'$:
\BY
&&\langle \h Y_{in}(t)\h Y_{in}(t')\rangle=\frac{1}{N}\sum_{n,n'=1}^N\langle \h Y(t-nT)\h Y(t'-n'T)\rangle.\L{35}
\EY
Figure \ref{CorrAfter} represents the spectrum of the photocurrent fluctuations  $S_{in}(\omega)$ (blue dashed line) under homodyne detection of the $ \h Y $ - quadrature of this signal:
\BY
&&S_{in}(\omega)=\frac{1}{2\pi}\int\int e^{i\omega t}e^{-i\omega t'}\beta(t')\beta(t)\langle \h Y_{in}(t)\h Y_{in}(t')\rangle dtdt'.\L{36}
\EY
For this calculation we have chosen the parameters of the system in agreement both with experimental studies of the SPOPO and the calculations presented above:
$N=90$ , $T_0=0.1$ , $T=10000$, $\kappa_sT=0.1$, $\Theta(t)=H(t)H(T_0-t)$. Time profile of the homodyne $\beta(t)$ has the shape of a step function and
coincides with the profile of the signal. In this case, the fluctuation spectrum is given by:
\BY
&& S_{in}(\omega)=1-\frac{\kappa_s T}{2N}\sum_{m,k=1}^N \cos((m-k)T\omega)e^{-\kappa_sT|m-k|}\;.\L{37}
\EY
The resulting spectrum shows the noise reduction at frequencies that are multiples of $2\pi/T$ within the spectral width of the comb.
Thus, we need to check whether dips are saved in the spectrum of the restored signal.

Another important feature of the quantum statistics of the SPOPO radiation required discussion is to find out the number of independent quantum degrees of freedom possessed by this light.
Let us remind that the comb consists of a large (about $ 10^5 $) number of frequency components, and each of them is formed in the process of parametric down-conversion,
so, at first glance, it may seem that this number characterizes the  number of  independent degrees of freedom of the system.
However, a detailed analysis \cite{Patera} showed that the number of true quantum degrees of freedom is not so high.
To characterize the quantum degrees of freedom, the authors introduced the concept of "supermodes"\, or in other words, of $q$-modes.
Let us briefly recall what it is.  The creation of the SPOPO’s photons in the process of parametric down-conversion of the pump is described by the effective Hamiltonian as follows:
\BY
&&\h{H}=i\hbar g\sum_{m,n}L_{m,n}\h a^\dag_m \h a^\dag_n+h.c.
\EY
Here the coupling constant $g$ is proportional to the amplitude of the pump and describes the resulting efficiency of the process, $a^\dag_m$ is the photon creation operator  associated with the mode of frequency
 $ \omega_m$, and  matrix $L_{m,n}$ describes the binding force between modes with frequencies $\omega_m$ and $\omega_n$. Matrix
$L_{m,n}$ is formed by two factors: this is the product of the phase matching  function $f_{m,n}$ and the spectral amplitude  $p_{m+n}$ of the pump at the frequency $\omega_m+\omega_n$:
\BY
&&L_{m,n}=f_{m,n}\cdot p_{m+n}\;.
\EY
The matrix $L_{m,n}$  is a huge matrix containing about $10^5 \times 10^5$ components. However, if we diagonalize and represent it as an expansion of the eigenvectors $\{X_k\}$,
\BY
L_{m,n}=\sum_{k}\Lambda_{k}X_{k,m}X_{k,n},
\EY
we can easily see that only a small number of the eigenvalues $\Lambda_{k}$ is nonzero.  Then it is possible to rewrite the Hamiltonian as follows:
\BY
\h{H}=i\hbar g\sum_{k}\Lambda_{k}\h{L}^{\dag 2}_k+h.c.,\qquad\mbox{where}\qquad \h{L}_k=\sum_{i}X_{k,i}\h{a}_i.\L{41}
\EY
Here operators $\h L_k$ are the supermodes, defined as linear combinations of the initial single-frequency modes.

Experimentally, the first six uncorrelated squeezed supermodes were observed  \cite{Roslund}.
We will follow the preservation of the squeezing precisely  in these supermodes to make sure that our memory scheme is able to preserve all correlations presented in the input signal.
Note that observation of six squeezed modes does not represent an inherent upper limit of
the quantum dimensionality of the comb states. Under the suitable improvements of the experimental setup (e.i., better adapting the pump spectrum,
increasing cavity bandwidths, etc.) states possessing as many as $\sim 100$ squeezed modes are expected \cite{Araujo}.

In  \cite{Patera}, it was demonstrated that profiles of the supermodes could be described  with good accuracy by the Hermite-Gaussian functions:
\BY
&& L_k(\omega)= \frac{1}{\sqrt{k! 2^k \sqrt{\pi} N_s}} \exp\(-\frac{1}{2}\(\frac{\omega-\omega_s}{N_s \omega_{rep}}\)^2\)\; H_k\(\frac{\omega-\omega_s}{N_s
\omega_{rep}}\),\L{42}
\EY
where $H_k$ is the Hermite polynomial of order $k$; $\omega_s$ is the carrier (central) frequency of the comb; $N_s = (\omega_{rep} T_0)^{-1}$ is the number of teeth in the comb; $\omega_{rep}$ is spectral interval between the modes.

Next, we will check whether the offered memory scheme preserves all quantum characteristics of the input signal that are mentioned above.

%-----------------------------------------------------------------------------

\section{Evaluation of the writing efficiency}\L{record}

Before proceeding with a discussion of the preservation of quantum correlations, we evaluate such a basic characteristic of the memory process as efficiency.
Let us see how the writing efficiency behaves depending on the number of pulses to be stored.

The  writing efficiency is defined as the ratio of the average number of spin excitations $N_{spin}$ to the total average number of photons in the input signal $N_{in}$ :
\BY
&&{\cal E}=\frac{N_{spin}}{N_{in}},\qquad N_{spin}=\int\limits_0^Ldz\;\langle \hat {b}^{W \dag}(z) \hat {b}^W(z)\rangle,\qquad
N_{in}=\int\limits_0^{T_W} dt\; \langle\hat a_{in}^\dag(t)\hat a_{in} (t)\rangle .
\EY
The total average number of excitations in the medium can be explicitly represented by the equation:
\BY
&&N_{spin}=\int\limits_0^Ldz\int\limits_{0}^{ T_W}\int\limits_{0}^{ T_W}
dt dt^\prime\;\langle\hat a^\dag_{in}(t)\hat a_{in}(t^\prime)\rangle
G_{ab}^*(t,z)G_{ab}(t^\prime,z).
\EY
Since to evaluate the efficiency we need only to know the average number of excitations in the input signal, here, we can assume all pulses to be the same, ignoring the fluctuations of the signal:
\BY
a_{in}(t)=aF(t)=\sum_{n=1}^N a_{n}(t)\Theta(t-t_n)=\sum_{n=1}^N \Theta(t-t_n).
\EY
Then, using the Schmidt decomposition, we obtain the following equation:
\BY
&&N_{spin}=\sum\limits_{i}\sqrt{\lambda_i}\;   | \int\limits_0^{T_W}dte^{i\frac{T_0}{T}t}\varphi_i(t)|^2.
\EY
Besides, under these conditions, the number of photons in the input signal is:
\BY
N_{in}=T_0N.
\EY
Therefore, the writing efficiency is given by:
\BY
&&{\cal E}=\frac{1}{ T_0N}\sum_{i}\sqrt{\lambda_i}\;| \int\limits_0^{T_W}dte^{i\frac{T_0}{T}t}\varphi_i(t)|^2.
\EY
Since the eigenfunctions in the Fourier representation $\varphi_i(\omega)$ can be derived as
\BY
&&\varphi_i(\omega)=\frac{1}{\sqrt{T_W}} \int\limits_0^{T_W} dt\;e^{i\frac{T_0}{T}t} \varphi_i(t)e^{ i\omega t},
\EY
then, taking into account the condition $T_0<<T$ and the approximate equality $T_W\simeq NT$, implemented under the condition  $N>>1$, we obtain the writing efficiency as:
\BY &&{\cal E}=\frac{T}{ T_0}\sum_{i}\sqrt{\lambda_i}\; |\varphi_i(\omega=0)|^2= \sum_{i}\sqrt{\lambda_i}\; \tilde{\varphi}^2_i(\omega=0) .\label{effy}
\EY

Let us turn again to Fig.\ref{eff}. As we already mentioned, the eigenvalues here are shown by the red bars, and the blue ones represent the values of the zero spectral components of the eigenfunctions.
Performing the summation in accordance with  Eq. (\ref{effy}), we obtain that the writing efficiency is about $0.9$. This indicates  that the work of the memory protocol under consideration (for given parameters) is satisfactory.

However, such a good result can be obtained not for all range of parameters.
It turns out, that the writing efficiency depends not only on the optical depth of the medium $L$ but also on the number of pulses in the train.

\begin{figure}[h]
 \begin{center}
  \includegraphics[height=50mm]{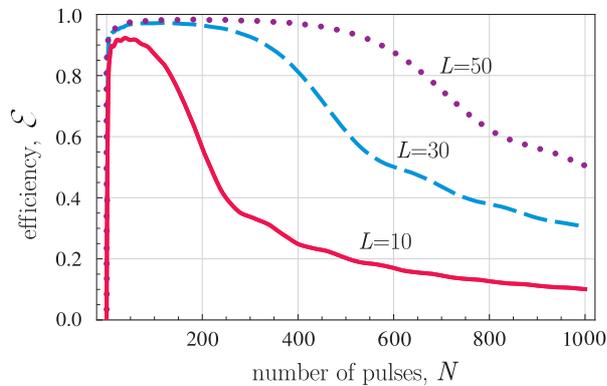}
 \caption{Dependence of the efficiency  of the recording stage on the number of pulses in a comb under the following parameters:
 $L=10, \;30, \;50$; $T_0=0.1$, $T=10000$.}
  \label{EffW}
 \end{center}
 \end{figure}

Figure \ref{EffW} shows the dependence of the writing efficiency on the number of pulses in the train at the fixed length of the medium and  duration of a single pulse.
As can be seen from the figure, the rapid growth of the efficiency replaces by the "plateau", followed by a decline.

This result is easy to understand, considering that the writing of each pulse of the train is accompanied by simultaneous readout and overwriting of the previously recorded field  on the deeper layers of the medium.
That is at the moment, when $n$\!--th pulse of the signal enters the medium, driving not only provides its record,
but also reads the information recorded by previous $(n-1)$ pulses, and, if possible, overwrites it in the deeper layers of the medium.
Thus, if the number of pulses becomes high, and the depth of the medium is not  large enough for rewriting, the driving "pushes"\,previously recorded information from the medium.
This result essentially distinguishes our problem from the problem in \cite{Nunn2007}, where it was considered the writing of a single short pulse on the three-level ensemble of atoms in the Raman configuration.
Due to the limitations, imposed on the durations of the processes,
such effects could not appear there as a matter of principle, and the writing efficiency tended to 1 if optical depth of the medium was large enough.

It is obvious that the reduction of the efficiency is formally  associated with the presence of the exponential factor in the kernel $K(t,t')$.
Here this factor spoils the result but still leaves the possibility to adjust proper parameters to enhance the efficiency, however, as we will show below, it plays a critical role in the preservation of the correlations.

%-----------------------------------------------------------------------------------

\section{Quantum correlations of the output signal}\L{reading}

As we have mentioned above, the sub-threshold radiation of the  SPOPO is characterized by the squeezing in the  $\h Y$-quadrature.
In accordance with (\ref{34})~-~(\ref{35}), normally ordered average of the  $Y$-quadrature correlator is:
\BY
&&\langle:\hat Y_{in}(t)\hat Y_{in}(t^\prime):\rangle=-\frac{\kappa_s T}{8N}\sum_{n,n'=1}^N e^{\ds -\kappa_s |t_n-t_{n'} |}
\delta(t-t^\prime-t_n+t_{n'})\Theta(t-t_n)\Theta(t^\prime-t_{n'}).
\EY
Let us follow how this correlator  will be transformed at the end  of the full cycle of the memory:
\BY
&&\langle:\h Y_{out}(t')\h Y_{out}(t''):\rangle=\int\limits_0^{T_W}\int\limits_0^{T_W}\cos \frac{T_0}{T}(t_1+t') \cos \frac{T_0}{T}(t_2+t'')
G(t_1,t')G(t_2,t'')\langle:\h Y_{in}(t_1)\h Y_{in}(t_2):\rangle dt_1dt_2\nn\\
&& +\int\limits_0^{T_W}\int\limits_0^{T_W}\sin \frac{T_0}{T}(t_1+t') \sin \frac{T_0}{T}(t_2+t'') G(t_1,t')G(t_2,t'')\langle:\h X_{in}(t_1)\h X_{in}(t_2):\rangle dt_1dt_2.
\EY
It is obvious that the presence of the second term, associated with the  stretched  quadrature  correlator  $\langle:~\h X_{in}(t_1)\h X_{in}(t_2):\rangle$ of the input signal, destroys the preservation of correlations in the restored field.
Despite that the multiplier $\frac{T_0}{T}$ in sine terms is small, it can not be neglected, because the time variable under the integral ranges from $ 0 $ to $ T_W $,
and therefore the argument of the trigonometric functions varies from $ 0 $ to  $ NT_0 $.
Moreover, this term can be dominant, since the value of the correlator of the stretched quadratures is considerably greater than of the squeezed ones.

To overcome this problem, we propose to modify the memory scheme by setting  two phase shifting devices at the input and output of the cell.
For example, acousto-optic modulator may serve as such a phase shifter.
Then the general design of the thought experiment is as follows (see Fig.  \ref{scheme_modified}):
before entering the memory cell, the SPOPO radiation passes through the phase shifter, that  linearly with time  changes the phase of the input signal to compensate for the phase factor $\exp (-it'\cdot T_0/T)$ (see Eq. (\ref{30})).
Then the light enters the memory cell, is written, then stored, and, after the readout, it goes once again to the phase shifter, which linearly with time deploys the phase of the output radiation in the opposite direction to compensate for the phase factor $\exp(it\cdot T_0/T)$.
\begin{figure}[h!]
\centering
\includegraphics[height=2.1cm]{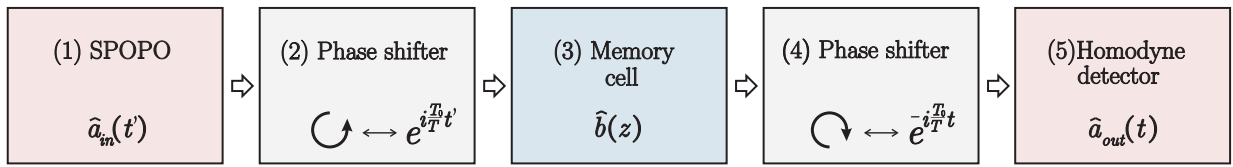}
\caption{A block diagram of the quantum memory with the phase shifters.} \L{scheme_modified}
\end{figure}
Then the kernel of integral transform in Eq. (\ref{30}) turns into symmetric and real kernel $G(t,t')$, and the $\h Y$-quadrature correlator of the restored signal depends only on the $\h Y$-quadrature correlator of the input one:
\BY
&&\langle:\h Y_{out}(t')\h Y_{out}(t''):\rangle=\int\limits_0^{T_W}\int\limits_0^{T_W}G(t_1,t')G(t_2,t'')\langle:Y_{in}(t_1)Y_{in}(t_2):\rangle dt_1dt_2.
\EY
Using the Schmidt decomposition (\ref{31}) and neglecting the changes of the eigenfunctions within the short time intervals of the duration $T_0$,  after the full memory cycle the correlator  can be expressed by the  eigenfunctions of the kernel $G(t,t')$:
\BY
&&\langle:\h Y_{out}(t')\h Y_{out}(t''):\rangle=-\frac{\kappa_s T}{8N}\sum\limits_{ij=1}\sqrt{\lambda_i\lambda_j};\ A_{ij}
\sum\limits_{m,k=1}^N\varphi_i(mT)\varphi_j(kT)\;\Theta(t'-mT)\Theta(t-kT),
\EY
where
\BY
A_{ij}=T_0\sum\limits_{m,k=1}^N e^{-\kappa_sT|k-m|}\varphi_i(mT)\varphi_j(kT).
\EY

Figure \ref{CorrAfter} shows the spectra $S_{out}(\omega)$ (red solid line) and $S_{in}(\omega)$ (blue dashed line)  of the shot noise suppression of the photocurrent of the $\h Y$-quadrature of the restored and input signals, respectively (see Appendix \ref{D}):
\BY
&&S_{out}(\omega)=1-\frac{\kappa_s TT_0}{2N}\sum_{m,k=1}^N \cos((m-k)T\omega)\sum_{i,j=1}\sqrt{\lambda_i\lambda_j}A_{ij}\varphi_i(mT)\varphi_j(kT),
\EY
\BY
&&S_{in}(\omega)=1-\frac{\kappa_s T}{2N}\sum_{m,k=1}^N \cos((m-k)T\omega)e^{-\kappa_sT|m-k|}.
\EY
The time profiles of the local oscillator in both spectra coincide with the profile of the driving. For the calculation we took into account the first six
eigenfunctions with the eigenvalues close to one. The resulting spectrum shows almost  the full preservation of the noise reduction at frequencies, which are multiples of $2\pi/T$ within the spectral width of the comb. Thus, we can say that our quantum memory scheme with the phase shifters allows one to preserve the quantum correlations of the input light with high efficiency.

Note that inclusion of the phase shifters also changes the efficiency of the scheme. Figure \ref{eff_saturation} shows the dependence of the writing efficiency on the number of pulses in the cases of the scheme with and without phase shifters.
\\
\begin{figure}[h!]
\centering
\includegraphics[height=5cm]{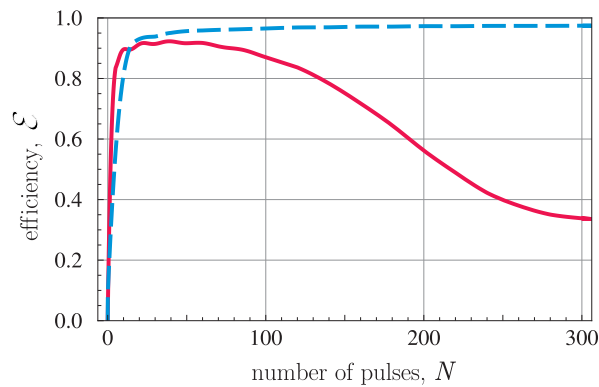}
\caption{The dependence of the writing efficiency on the number of pulses in the scheme with (blue dashed line) and without (red solid line) the phase shifters. Parameters of the calculation: $L=10$, $T_0=0.1$, $T=10000$.}
\L{eff_saturation}
\end{figure}
\\
It can be seen that for the train durations $NT$ under consideration, the decline of the efficiency does not occur. Note that this approach is correct for $N<1000$, since here the dimension time of  interaction of the signal  with the medium $ NT_0 $  should be less than the correspondent  length of the atomic ensemble  $ L / c $.

%--------------------------------------------------------------------------------------

\section{ Preservation of the squeezing in the supermodes}\L{VII}

In this section, we will follow how the first six supermodes of the SPOPO, observed in the experiment \cite{Roslund}, are stored in the memory cell.
In particular, we want to find out how well the model of quantum memory under study can preserve the squeezing in each individual  supermode.
We express the SPOPO field that goes to the input of the memory cell, as an expansion by a full orthonormal set of the functions $L_k(t)$, representing time profiles of the corresponding supermodes in Eq. (\ref{41}):
\BY
&& \hat E_{in} (t) = \sum_k  L_k(t) \hat e_{k}. \L{7_1}
\EY
Here $\h e_k$ is  annihilation operator of the photon in the $ k $\!--th supermode  that obeys the standard commutation relations:
\BY
[\h e_k, \h e_{k^\prime}^\dag]=\delta_{k,k^\prime},
\EY
and is normalized so that value $<\h e^\dag_k\h e_k>$ gives the number of photons in the $k$\!--th supermode.

Remind that time-dependent Hermite-Gaussian  functions  $L_k(t)$ satisfy the completeness and orthonormality conditions:
\BY
\sum_k L_k(t)L_k(t^\prime)=\delta(t-t^\prime),\qquad \int dt L_k(t) L_{k^\prime}(t) = \delta_{k, k^\prime}.
\EY
We consider the experimental scheme with the phase shifters as described in Section \ref{reading} (Fig.\ref{scheme_modified}), so, according to  Eq. (\ref{30}), we can derive the field at the output of the memory cell via the Schmidt modes (\ref{31}):
\BY
&& \hat E_{out} (t) = \sum_i \hat e_{i}\sum_{j}\sqrt{\lambda_i}C_{ij}\widetilde{\varphi_i}(t),\qquad C_{ij}=
\int\limits_0^{T_W}\;dt\widetilde{\varphi_i}(t)L_j(t), \L{7_2}
\EY
where $C_{ij}$ is  overlap integral for  the $i$\!--th Schmidt mode with $j$\!--th supermode.

Let us check how the squeezing of each specific supermode $k$ ($k=1,...,6$) is preserved by calculating  the variance of each mode at the input and output of the memory cell:
\BY
&& \langle:\big|\hat i_{\{in,out\},k} (\omega)\big|^2:\rangle =
\frac{1}{2\pi}\int\limits_{-\infty}^{+\infty}\int\limits_{-\infty}^{+\infty}\;dtdt'\langle: \hat i_{\{in,out\},k} (t)\hat i_{\{in,out\},k} (t')
:\rangle e^{i\omega(t-t')}.
\EY
For this aim we derive the expression of the photocurrent under homodyne detection, where the profile of the local oscillator coincides with the profile of the detected supermode:
\BY
&& \hat i_{in,k} (t) \sim L_k(t) e^{i\phi_{LO}} L_k(t) \hat e^\dag_{k} + h.c.,\\ \L{7_3}
&& \hat i_{out,k} (t) \sim L_k(t) e^{i\phi_{LO}} \hat e^\dag_{k}\sum_{i}\sqrt{\lambda_i}C_{ik}\widetilde{\varphi_i}(t) + h.c. \L{7_4}
\EY
Here $\phi_{LO}$ is the phase of the local oscillator chosen by the following considerations.
As it was shown in \cite{Patera,Roslund}, squeezed quadrature of the SPOPO supermodes alternate, i.e.,
 if $\h X$-quadrature is squeezed in $ n $\!--th supermode, then for $(n+1)$\!--th supermode it is $\h Y$-quadrature to be squeezed.
Thus, the local oscillator phase $\phi_{LO}$ should to be chosen so that its sum with the field phase $\phi_k$ would be equal to either $0$ for $\h X$-quadrature detection,  or $\pi/2$  for the detection of the $\h Y$-quadrature:
\BY
&& \hat i_{in,k} (t) \sim  L_k(t)^2 \hat X_{k},\qquad \hat i_{out,k} (t) \sim L_k(t) \hat X_{k}\sum_{i}\sqrt{\lambda_i}C_{ik}\widetilde{\varphi_i}(t),\qquad (\phi_{LO}+\phi_k=0) \L{7_5} \\
&& \hat i_{in,k} (t) \sim L_k(t)^2 \hat Y_{k},\qquad \hat i_{out,k} (t) \sim L_k(t) \hat Y_{k}\sum_{i}\sqrt{\lambda_i}C_{ik}\widetilde{\varphi_i}(t).\qquad (\phi_{LO}+\phi_k=\pi/2) \L{7_6}
\EY
Since the highest level of squeezing is achieved at zero frequency \cite{Patera}, we consider the variances at this value.
We make Fourier transformation of Eqs. (\ref{7_5})~-~(\ref{7_6}), put  $\omega=0$, and, taking into account the orthonormality of the  functions $L_k(t)$,
get the variances to compare the squeezings of the  $k$\!--th supermode at the  input and output of the memory cell:
\BY
&& \langle:\big|\hat i_{in,k} (\omega=0)\big|^2:\rangle \sim  \langle:\big| \hat X_{k}\big|^2:\rangle,\;\;\; \hat i_{out,k} (\omega=0) \sim
\langle:\big| \hat X_{k}\big|^2:\rangle(\sum_{i}\sqrt{\lambda_i}C_{ik}^2)^2,\;\;\; (\phi_{LO}+\phi_k=0)\L{7_8}\\ 
&&  \langle:\big|\hat i_{in,k} (\omega=0)\big|^2:\rangle \sim  \langle:\big| \hat Y_{k}\big|^2:\rangle,\;\;\; \hat i_{out,k} (\omega=0) \sim
\langle:\big| \hat Y_{k}\big|^2:\rangle(\sum_{i}\sqrt{\lambda_i}C_{ik}^2)^2.\;\;\; (\phi_{LO}+\phi_k=\pi/2) \L{7_9}
\EY
These expressions show that the preservation of the squeezing in the supermodes is determined by the value of the factor $\sum_{i}\sqrt{\lambda_i}C_{ik}^2$,
i.e., how the SPOPO‘s mode structure, defined by the parametric transformation in the crystal of source, corresponds to the mode structure  of the memory  cell.
\begin{figure}[h!]
\centering
\includegraphics[height=5.5cm]{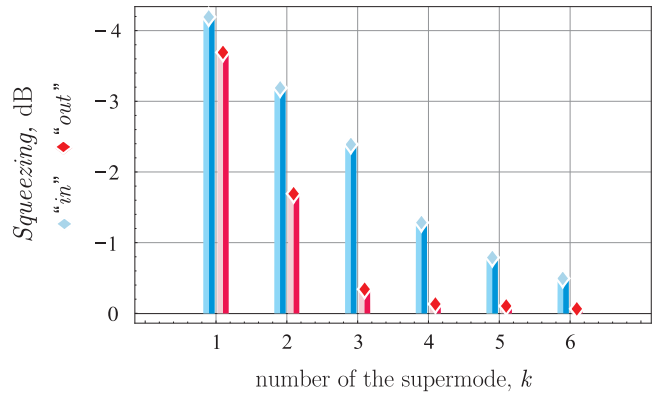}
\caption{Preservation of the squeezing in the first six supermodes. Values of the maximum squeezing in dB for SPOPO light before (blue bars, the values are taken from the experiment \cite{Roslund}) and after (red bars, the values are calculated in accordance with (\ref{7_8})~-~(\ref{7_9}))  the full cycle of the memory.
Parameters of the calculation: $L=10$, $T_0=0.1$, $T=10000$.}
\L{squeezing}
\end{figure}
Figure \ref{squeezing} demonstrates the values of the squeezing corresponding to the quadratures of the first six supermodes of the SPOPO before reaching the memory cell
(blue bars, the values are taken from the experiment \cite{Roslund}) and after it (red bars, the values are calculated in accordance with (\ref{7_8})~-~(\ref{7_9})). 
One can see that the first ($ -3.7 $ dB), second ($ -1.7 $ dB )and third ($ -0.4 $ dB) supermodes are squeezed significantly in the restored light.
For the rest of modes squeezing is about $ -0.1 $ dB, nevertheless it is still below the standard quantum limit.
This means that only three in six quantum degrees of freedom providing by SPOPO, i.e.,  independent squeezed orthogonal modes, remain after the memory transformation.
However, to give a fair assessment of these results, we need to make some important remarks.

First of all, initially (before the memory cell) the supermodes represented here were squeezed differently, and for the first three of them the squeezing was highly significant (less than $ -2 $ DB), 
while for the latter it was about $ -1 $ dB. As shown in \cite{Patera},
the experimental values of the squeezing of each supermode are far from the values of the possible maximum predicted by the theory.
The paper provides recommendations how to essentially improve the initial squeezing and to make it about $ -5 $ dB, that, evidently, will improve the output squeezing.

On the other hand, the chosen procedure of the homodyne detection for restored light is not optimal, since it does not take into account the distortion caused by the memory (\ref{30}).
We does not  change the time profile of the local oscillator, because we are concerned about how the squeezing is preserved in the selected basis of the supermodes, i.e., in a field with a given time profile.
Note that the distortions are caused by the fact that with the given parameters, there are only six Schmidt modes $\varphi_i(t)$ with $\sqrt{\lambda_i}\approx1$ that one can use to decompose  the supermodes  in Eq. (\ref{7_2}).
Accordingly, we can reduce the influence of this factor by increasing the optical depth of the memory cell, thereby increasing the number of the Schmidt modes with $\sqrt{\lambda_i}\approx1$.

Finally, as it was shown in \cite{Gorshkov}, to preserve the modes of the field with a particular profile, one can utilize the method of the optimization based on adjusting the profile of the driving.
Remind that up until now we have considered the simplest shape of the driving field with a rectangular profile, so there is another parameter to vary.

Thus, we can conclude that even without additional manipulations, we are able to preserve the first three squeezed supermodes of the SPOPO in one memory cell and, if necessary, to increase their number to six,
i.e. to preserve all six supermodes obtained in the experiment \cite{Roslund}.

%--------------------------------------------------------------------------------------

\section{ Conclusion}

In this paper, we demonstrated that the offered memory scheme allows one to preserve the SPOPO light correlations at the quantum level, including
the genuine multipartite entanglement embedded in this light.

We followed the signal writing efficiency and showed that for the chosen parameters of the SPOPO radiation, which correspond to experimentally obtained values and for the  dimensionless optical depth $L=10$,
it is possible to effectively preserve the train of $\sim 100$  pulses.
The phase correction technic allows one to increase this number.
Although the use of the phase shifters in the memory scheme is not a crucial requirement to obtain the high efficiency of its operation,
we have shown that without these elements the preservation of the quadrature squeezing is impossible.
Without the phase shifters  the  phase and amplitude quadratures of the  signal evolve not independently,
and the fluctuation spectrum of the restored field would be determined by both the squeezed and stretched quadratures of the input signal.
Note that the phase shifting may be realized by different devices, for example, the phase shifting of the output signal can be easily replaced by choosing the appropriate phase of the homodyne.
However, such a recipe would not work at the input of the memory cell. 
Analysis of the solutions shows that one should change the design of light-matter interaction (the kernel of the interaction, formally) by compensating for the phase shifts purchased by the pulses during propagation.

An important part of the work was devoted to estimate the number of the independent quantum degrees of freedom of the restored radiation. To do this we applied the technique of the supermodes developed in the series of the studies
 \cite{Patera,Roslund,Araujo,Gerke}. We have compared the squeezings of the supermodes of the signal at the input and the output of the memory cell, taking the experimentally obtained data as the input parameters.
Our calculation has shown that even at a relatively low optical depth, all six  squeezed supermodes of the input signal, would be squeezed in the restored radiation, but only the first three would have the essential squeezing.
This allows us to assert that the frequency entanglement presented in the recovered comb can be arranged in a way to embed multiple cluster states in its structure, as it is in the case of the input radiation of SPOPO. 
So, we are able to fabricate the cluster states on this basis in the future and manipulate them inside as well as outside the memory cell.
\\

The reported study was supported by RFBR (Grants 15-02-03656a, 16-02-00180a and 16-32-00594).

%-------------------------------------------------------------------------------

%---------------------------------------------------------------------------
\appendix
\section{Simplifying expressions of the spin coherences}\L{A}

Let us show that we can put  $B=0$ in the exponent in Eq. (\ref{17}) for the spin coherence. Remember that initially the problem was formulated for the fast varying amplitudes. Let us replace them back in the obtained solution:
\BY
&&\hat {b}^W(z)\to \hat {b}^W(z)e^{ i(k_d-k_s)z}.
\EY
Then Eq. (\ref{17}) can be derived as:
\BY
&&[\hat {b}^W(z)\;e^{i(k_d-k_s)z}]=\\
&& =-iCe^{ i (k_d - B)z}\sum_{n=1}^N\int\limits_{t_n}^{t_n+T_0} dt\;[\hat a_{in}(t)e^{- ik_sz}] e^{-iA(D_n+t_n-t)}
\;J_0\(2C\sqrt{(D_n+t_n-t)z}\)+vac.\nn
\EY
One can see, the value $ B $ is an additive to the wave number of the driving $ k_d $. It is easy to be convinced that for  the discussed parameters of the problem this correction is negligibly small.
By setting   $\Delta\simeq 10^7\gamma$, $d \lesssim 100$,
$L/\lambda\simeq 1000$ as estimates, we get:
\BY
&&\frac{k_d}{B}=\frac{k_d\Delta}{g^2N_{at}}=\frac{\Delta}{\gamma}\; \frac{1}{d}\;\frac{L}{\lambda}=10^7\;\frac{1}{100}\;1000\gg 1.
\EY
%

%---------------------------------------------------------------------------
\section{Relation between the kernels of integral transforms for the  writing and the readout}\L{B}

Let us show that the kernel  $G_{ab}(t,z)$,  describing the writing, and the kernel $G_{ba}(t,z)$, related to the reading, are interconnected as follows:
\BY
G_{ab}(T_W-t,z)=G_{ba}(t,z),
\EY
where
$$T_W=(N-1)T+T_0.$$
To prove this equality, we derive the initial expressions of the kernels to be compared:
\BY
&&G_{ab}(t,z)=\sum\limits_{k=1}^{N}\Theta(t-t_k)e^{ -i((N-k+1)T_0+t_k-t)}J_0\(2\sqrt{z((N-k+1)T_0+t_k-t)}\),\\
&&G_{ba}(t,z)=\sum\limits_{n=1}^{N}\Theta(t-t_n)e^{ -i((n-1)T_0+t-t_n)} J_0\(2\sqrt{z((n-1)T_0+t-t_n)}\),
\EY
where $t_i=(i-1)T$   ($i=n,k$), and $\Theta(t)=H(t)\cdot H(T_0-t)$ is a window-function with duration $T_0$.\\
\\
Then,
\BY
&&G_{ab}(T_W-t,z)=G_{ab}(-t+(N-1)T+T_0,z)\nn\\
&&=\sum\limits_{k=1}^{N}\Theta(T_0-t+(N-k)T)e^{i(  (N-k)T_0+t+(k-N)T   )}J_0\(2\sqrt{z((N-k)T_0+t+(k-N)T)}\)\nn\\
&&[k=N-n+1]\nn\\
&&=\sum\limits_{n=1}^{N}\Theta(T_0-t+(n-1)T)e^{i((n-1)T_0+t-(n-1)T)}J_0\(2\sqrt{z((n-1)T_0+t-(n-1)T))}\)\nn\\
&&=\sum\limits_{n=1}^{N}H(T_0-t+(n-1)T)\cdot H(T_0-T_0+t-(n-1)T) \cdot e^{i((n-1)T_0+t-t_n)}J_0\(2\sqrt{z((n-1)T_0+t-t_n))}\)\nn\\
&&=\sum\limits_{n=1}^{N}\Theta(t-t_n)e^{i((n-1)T_0+t-t_n)}J_0\(2\sqrt{z((n-1)T_0+t-t_n))}\)=G_{ba}(t,z).\nn
\EY

%---------------------------------------------------------------------------
\section{Approximation of the function $Q(t)$} \L{C}

\begin{figure}[h]
 \begin{center}
  \includegraphics[height=60mm]{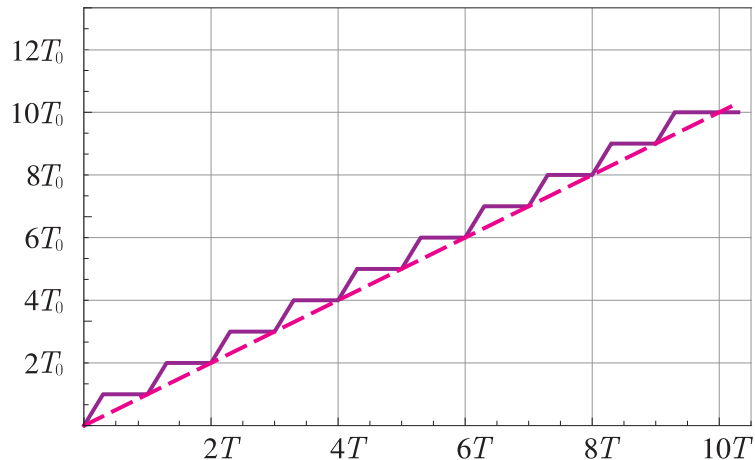}
 \caption{ Functions $Q(t)$ (purple solid line) and its approximating straight line $\frac{T_0}{T}t$  (pink dashed line). Parameters of the calculation: $T_0=3$, $T=10$. }
  \label{Approx}
 \end{center}
 \end{figure}

Let us illustrate that under a condition of a high number of pulses
in the train ($N>>1$), function $Q(t)$ can be approximated by a straight line:
\BY
&& Q(0,t)=\int\limits_{0}^{t}dt'F(t')\simeq\frac{T_0}{T}t.
\EY
Figure  \ref{Approx} represents the function $Q(t)$ and the line $\frac{T_0}{T}t$, calculated for the same parameters ($T=10$, $T_0=3$).
Here we have chosen  the values $T$ and $T_0$ to be of  a same order to visualize the step-form of the function $Q(t)$ on the graph. 
With increasing of the number of pulses in the train the number of "steps"\,also increases, but the shape of the function does not change, 
and it can be clearly seen, the greater is the number of pulses the less important is the fine structure of $ Q (t) $ in comparison with its pronounced linear growth, so the approximation improves. 
The ratio $T_0/T$ corresponds to the slope of the approximating straight line.
Note that by approaching the parameters of calculation to real ones, the slope of the line becomes close to zero, that does not affect the possibility of approximation, but makes its visualization  unclear.

%---------------------------------------------------------------------------
\section{Spectra of the photocurrent fluctuations under homodyne detection} \L{D}

Let us derive in the explicit form the spectra of photocurrent fluctuations under homodyne detection of  $\h Y$-quadrature correlation functions  before and after the full memory cycle. According to (\ref{34})-(\ref{35}), the correlation function for $\h Y$-quadratures of SPOPO light is given by:
\BY
&&\langle \h Y_{in}(t)\h Y_{in}(t')\rangle=\frac{1}{N}\sum_{n,n'=1}^N\langle \h Y(t-nT)\h Y(t'-n'T)\rangle\nn\\
&&=\frac{1}{4}\frac{1}{N}\sum_{n,n'=1}^N\delta_{nn'}\delta(t-t')-\frac{\kappa_sT}{8N}\sum_{n,n'=1}^Ne^{-\kappa_sT|n-n'|}\;\delta(t-t'-(n-n')T)\Theta(t-nT)\Theta(t'-n'T).
\EY
To detect this correlation function, the homodyne detection is applied, where the profile $\beta(t)$ of the local oscillator is represented by the train of the rectangular pulses:
\BY
 \beta(t)=\sum_{n=1}^N\Theta(t-t_n),\qquad t_n=(n-1)T. \label{D2}
\EY
Thus, the expression of the photocurrent spectrum under homodyne detection is as follows:
\BY
&& \langle|i_{in}(\omega)|^2\rangle=\langle i_{in}(\omega)i_{in}(-\omega)\rangle\nn\\
&&=\frac{1}{2\pi}\int\int e^{i\omega t}e^{-i\omega t'}\beta(t')\beta(t)\langle \h Y_{in}(t)\h Y_{in}(t')\rangle dtdt'\nn\\
&&=\frac{N}{4N}\frac{T_0}{2\pi}-\frac{\kappa_sT}{8N}\sum_{n,n'=1}^N\frac{1}{2\pi}\int\int e^{i\omega t}
e^{-i\omega t'}e^{-\kappa_sT|n-n'|}\;\delta(t-t'-(n-n')T)\Theta(t-nT)\Theta(t'-n'T)dtdt'\nn\\
&&=\frac{1}{4}\frac{T_0}{2\pi}-\frac{\kappa_sT}{8N}\sum_{n,n'=1}^N\frac{1}{2\pi}\int\limits_{nT}^{nT+T_0}\int\limits_{n'T}^{n'T+T_0}
e^{i\omega t}e^{-i\omega t'}e^{-\kappa_sT|n-n'|}\;\delta(t-t'-(n-n')T)dtdt'\nn\\
&&=\frac{1}{4}\frac{T_0}{2\pi}-\frac{\kappa_sT}{8N}\frac{1}{2\pi}\sum_{n,n'=1}^Ne^{-\kappa_sT|n-n'|}\int\limits_{nT}^{nT+T_0}
e^{i\omega t}e^{-i\omega(t-(n-n')T)}dt\nn\\
&&=\frac{1}{4}\frac{T_0}{2\pi}-\frac{\kappa_sT}{8N}\frac{T_0}{2\pi}\sum_{n,n'=1}^Ne^{-\kappa_sT|n-n'|}e^{i\omega T(n-n')}\nn\\
&&=\frac{1}{4}\frac{T_0}{2\pi}-\frac{\kappa_sT}{8N}\frac{T_0}{2\pi}\sum_{n,n'=1}^Ne^{-\kappa_sT|n-n'|}cos\(\omega T(n-n')\).
\EY
Dividing this expression by the square of the shot noise photocurrent, we finally obtain a normalized spectrum of  photocurrent fluctuations at the input of the memory cell:
\BY
&& S_{in}(\omega)=1-\frac{\kappa_s T}{2N}\sum_{n,n'=1}^N \cos((n-n')T\omega)e^{-\kappa_sT|n-n'|}.\L{E4}
\EY
Now let us consider the spectrum of the output (restored) signal. The first term in Eq. (\ref{34}) describes the shot noise and remains unchanged during the memory process. 
The second one corresponds to the normally ordered correlator and is responsible for the degree of the shot noise  reduction.   After the full memory cycle it can be derived with the help of Schmidt modes:
\BY
&&\langle:\h Y_{out}(t')\h Y_{out}(t''):\rangle=\int\limits_0^{T_W}\int\limits_0^{T_W}G(t_1,t')G(t_2,t'')\langle:\h Y_{in}(t_1)\h Y_{in}(t_2):\rangle dt_1dt_2\nn\\
&&=\int\limits_0^{T_W}\int\limits_0^{T_W}\sum_{ij}\sqrt{\lambda_i\lambda_j}\varphi_i(t_1)\varphi_i(t')\varphi_j(t_2)\varphi_j(t'')
\langle:\h Y_{in}(t_1)\h Y_{in}(t_2):\rangle dt_1dt_2\nn\\
&&=-\sum_{ij}\sum_{n,n'=1}^N\sqrt{\lambda_i\lambda_j}\int\limits_0^{T_W}\int\limits_0^{T_W}\varphi_i(t_1)\varphi_i(t')\varphi_j(t_2)\varphi_j(t'')\nn\\
&&\times\frac{\kappa_s T}{8N}\; e^{\ds -\kappa_sT |n-n' |}\Theta(t_1-nT)\Theta(t_2-n'T)\delta\(t-t'-(n-n')T\)dt_1dt_2\nn\\
&&=-\frac{\kappa_s T}{8N}\sum_{ij}\sum_{n,n'=1}^N\sqrt{\lambda_i\lambda_j} \; e^{\ds -\kappa_sT |n-n' |}\varphi_i(t')\varphi_j(t'')\nn\\
&&\times\int\limits_{nT}^{nT+T_0}\int\limits_{n'T}^{n'T+T_0}\varphi_i(nT)\varphi_j(n'T)\delta\(t-t'-(n-n')T\)\;dt_1dt_2\nn\\
&&=-\frac{\kappa_s T}{8N}\sum_{ij}\sum_{n,n'=1}^N\sqrt{\lambda_i\lambda_j} \; e^{\ds -\kappa_sT |n-n' |}
\varphi_i(t')\varphi_j(t'')\varphi_i(nT)\varphi_j(n'T)\int\limits_{nT}^{nT+T_0}\;dt_1\nn\\
&&=-\frac{\kappa_s T}{8N}T_0\sum_{ij}\sum_{n,n'=1}^N\sqrt{\lambda_i\lambda_j} \; e^{\ds -\kappa_sT |n-n' |}
\varphi_i(t')\varphi_j(t'')\varphi_i(nT)\varphi_j(n'T)\nn\\
&&=-\frac{\kappa_s T}{8N}\sum_{ij}\sqrt{\lambda_i\lambda_j} \; A_{ij}\;\varphi_i(t')\varphi_j(t''),
\EY
where
\BY
A_{ij}=T_0\sum\limits_{n,n'=1}^Ne^{-\kappa_sT|n-n'|}\varphi_i(nT)\varphi_j(n'T).
\EY
Neglecting the evolution of the eigenfunctions within short time intervals of duration $ T_0 $, we can directly derive the correlator after the full memory cycle over the mean values of the eigenfunctions  on these intervals:
\BY
&& \langle:\h Y_{out}(t')\h Y_{out}(t''):\rangle=-\frac{\kappa_s
T}{8N}\sum\limits_{ij=1}\sqrt{\lambda_i\lambda_j}A_{ij}\sum\limits_{m,k=1}^N\varphi_i(mT)\varphi_j(kT)\;\Theta(t'-mT)\Theta(t''-kT).
\EY
Applying the homodyne detection technic with the same local oscillator profile (\ref{D2}), we obtain:
\BY
&& \langle|i_{out}(\omega)|^2\rangle=\frac{1}{4}\frac{T_0}{2\pi}-\frac{\kappa_s T}{8N}\frac{T_0^2}{2\pi}\sum_{m,k=1}^N
\cos((m-k)T\omega)\sum_{i,j=1}\sqrt{\lambda_i\lambda_j}A_{ij}\varphi_i(mT)\varphi_j(kT).
\EY
Similar to the previous consideration, we divide the resulting expression by the square of the shot noise intensity, and get a normalized spectrum of the photocurrent fluctuations of the restored signal  after the full  memory cycle:
\BY
&& S_{out}(\omega)=1-\frac{\kappa_s TT_0}{2N}\sum_{m,k=1}^N \cos((m-k)T\omega)\sum_{i,j=1}\sqrt{\lambda_i\lambda_j}A_{ij}\varphi_i(mT)\varphi_j(kT).
\EY

\end{document}